\documentclass[a4,10pt]{article}

\usepackage{amsmath}
\usepackage{amssymb}
\usepackage{indentfirst}

\newtheorem{theorem}{Theorem}
\newtheorem{lemma}{Lemma}

\renewcommand{\epsilon}{\varepsilon}
\renewcommand{\phi}{\varphi}

\newcommand{\Def}{\ensuremath{\stackrel{\rm def}=}} 

\newcommand{\ST}{\ensuremath{\quad\text{such that}\quad}}
\newcommand{\AND}{\ensuremath{\quad\text{and}\quad}}

\renewcommand{\H}{\ensuremath{{\cal H}}}
\renewcommand{\S}{\ensuremath{{\cal S}}}

\renewcommand{\L}{\ensuremath{{\cal L}}}

\newcommand{\E}{\ensuremath{{\cal E}}}

\newcommand{\Hn}{\ensuremath{{\cal H}^{\otimes n}}}
\newcommand{\rhon}{\ensuremath{\rho^{\otimes n}}}
\newcommand{\sigman}{\ensuremath{\sigma^{\otimes n}}}

\newcommand{\Tr}{\ensuremath{\mbox{\rm Tr}}}

\newcommand{\proj}[1]{\ensuremath{\left\{ #1 \right\}}}
\newcommand{\oneover}[1]{\ensuremath{\frac{1}{#1}}}
\newcommand{\defset}[2]{\ensuremath{%
 \left\{#1\,\left|\,#2\right.\right\}
}}

\newcommand{\pinching}[2]{\ensuremath{\,\E_{#1}\!\!\left(#2\right)}}

\newcommand{\bigzeroUR}{{\lower0.8ex\hbox{\Large 0}}}
\newcommand{\bigzeroDL}{{\raise0.8ex\hbox{\Large 0}}}

\def\appendix{\par
    \setcounter{section}{0}\setcounter{subsection}{0}
    \def\thesection{\Alph{section}} \section*{Appendix}
}
\def\appendices{\par
    \setcounter{section}{0}\setcounter{subsection}{0}
    \def\thesection{\Alph{section}} \section*{Appendices}
}

\begin{document}

\title{
A New Proof of the Direct Part of Stein's Lemma\\
in Quantum Hypothesis Testing
}

\author{
Tomohiro Ogawa\thanks{
Department of Mathematical Informatics,
Graduate School of Information Science and Technology,
The University of Tokyo,
7-3-1 Hongo, Bunkyo-ku, Tokyo, 113-0033 Japan.
(e-mail: ogawa@sr3.t.u-tokyo.ac.jp)
}
\and
Masahito Hayashi\thanks{
Laboratory for Mathematical Neuroscience, Brain Science Institute, RIKEN,
2-1 Hirosawa, Wako, Saitama, 351-0198 Japan.
(e-mail: masahito@brain.riken.go.jp)
}
}

\date{}

\maketitle

\noindent{\bf Abstract}---
The direct part of Stein's lemma in quantum hypothesis testing
is revisited based on a key operator inequality
between a density operator and its pinching.
The operator inequality is used to show
a simple proof of the direct part of Stein's lemma
without using Hiai-Petz's theorem,
along with an operator monotone function,
and in addition
it is also used to show a new proof of Hiai-Petz's theorem.


\section{Introduction}

Quantum hypothesis testing is a fundamental problem
in quantum information theory,
because it is one of the most simple problem
where the difficulty derived from
noncommutativity of operators appears.
It is also closely related to other topics in quantum
information theory, as in classical information theory.
Actually, its relation with quantum channel coding is discussed
in \cite{Ogawa-Nagaoka-2001} \cite{Hayashi-Nagaoka}.

Let $\H$ be a Hilbert space which represents a physical system in interest.
We assume $d\Def\dim\H<\infty$ for mathematical simplicity.
Let $\L(\H)$ be the set of linear operators on $\H$
and define the set of density operators on $\H$ by 
\begin{align}
\S(\H) \Def \defset{\rho\in\L(\H)}{\rho=\rho^*\ge 0, \Tr[\rho]=1}.
\end{align}
We study the hypothesis testing problem for
the null hypothesis $H_0 : \rho_n\Def\rhon\in\S(\Hn)$
versus the alternative hypothesis $H_1 : \sigma_n\Def\sigman\in\S(\Hn)$,
where $\rhon$ and $\sigman$ are the $n$th
tensor powers of arbitrarily given density operators 
$\rho$ and $\sigma$ in $\S(\H)$.

The problem is to decide which hypothesis is true
based on the data drawn from a quantum measurement,
which is described by a positive operator valued measure (POVM)
on $\Hn$, i.e.,
a resolution of identity $\sum_i M_{n,i} = I_n$
by nonnegative operators $M_n=\{M_{n,i}\}$ on $\Hn$.
If a POVM consists of projections on $\Hn$,
it is called a projection valued measure (PVM).
In the hypothesis testing problem, however,
it is sufficient to treat a two-valued POVM $\{M_0,M_1\}$,
where the subscripts $0$ and $1$ indicate the acceptance
of $H_0$ and $H_1$, respectively.
Thus, an operator $A_n\in\L(\Hn)$ satisfying inequalities
$0\le A_n\le I_n$ is called a test
in the sequel, since $A_n$ is 
identified with the POVM $\{A_n,\, I_n-A_n\}$. 
For a test $A_n$, the error probabilities of the first kind and 
the second kind are, respectively, defined by
\begin{align}
\alpha_n (A_n) \Def \Tr[\rho_n (I_n -A_n)], \quad
\beta_n (A_n) \Def \Tr[\sigma_n A_n].
\end{align}

Let us define
\begin{align}
\beta_n^*(\epsilon)\Def\min
\bigl\{ \beta_n(A_n) &\bigm| A_n:\text{test},\,
\alpha_n(A_n)\le\epsilon \bigr\},
\end{align}
and the quantum relative entropy:
\begin{align}
D(\rho\|\sigma)\Def\Tr[\rho(\log\rho-\log\sigma)].
\end{align}
Then we have the following theorem, which is one of the most
essential theorem in quantum information theory.
\begin{theorem}[Stein's lemma]
\begin{align}
\lim_{n\rightarrow\infty}\oneover{n}\log\beta_n^*(\epsilon)=-D(\rho\|\sigma).
\label{Stein}
\end{align}
\end{theorem}
The first proof of \eqref{Stein} was composed of two inequalities.
One is the direct part given by Hiai-Petz \cite{Hiai-Petz}:
\begin{align}
\limsup_{n\rightarrow\infty}\oneover{n}\log\beta_n^*(\epsilon)
\le -D(\rho\|\sigma),
\label{direct1}
\end{align}
which takes an equivalent form (see \cite{Ogawa-Nagaoka-2000}):
\begin{align}
&\exists \{A_n:\text{test}\}_{n=1}^{\infty}
\ST
\lim_{n\rightarrow\infty}\alpha_n(A_n)=0
\AND
\limsup_{n\rightarrow\infty}\oneover{n}\log\beta_n(A_n)
\le -D(\rho\|\sigma),
\label{direct2}
\end{align}
and the other is the converse part
given by Ogawa-Nagaoka \cite{Ogawa-Nagaoka-2000}.
A direct proof of the equality \eqref{Stein}
was also given by Hayashi \cite{Hayashi-Stein}
using the information spectrum approach in quantum setting \cite{Nagaoka}.

Preceding the direct part \eqref{direct1},
Hiai-Petz \cite{Hiai-Petz} proved an important theorem
which is explained as follows.
Let $D_{M_n}(\rho_n\|\sigma_n)$
be the classical relative entropy (Kullback divergence)
between the probability distributions $\bigl\{\Tr[\rho_n M_{n,i}]\bigr\}$
and  $\bigl\{\Tr[\sigma_n M_{n,i}]\bigr\}$.
Then the monotonicity \cite{Lindblad-CP} \cite{Uhlmann}
of the quantum relative entropy yields
\begin{align}
D(\rho\|\sigma)\ge\oneover{n}D_{M_n}(\rho_n\|\sigma_n)
\label{monotonicity1}
\end{align}
for any POVM $M_n$,
and there exists a POVM that attains
the equality if and only if $\rho$ and $\sigma$ mutually commute.
In other words, the right-hand side (RHS) of
\eqref{monotonicity1} is less than the quantum relative entropy
for any POVM in general.
In this situation, however,
Hiai-Petz proved that the RHS of
\eqref{monotonicity1} with good POVMs
can achieve the quantum relative entropy asymptotically
as follows.
\begin{theorem}[Hiai-Petz \cite{Hiai-Petz}]
\begin{align}
D(\rho\|\sigma)
= \lim_{n\rightarrow\infty} \oneover{n}
\sup_{M_n} D_{M_n}(\rho_n\|\sigma_n),
\label{Hiai-Petz-theorem}
\end{align}
where the supremum is taken over the set of POVMs on $\Hn$.
\end{theorem}
\vspace{1ex}
They combined \eqref{Hiai-Petz-theorem} with the classical hypothesis
testing problem to show the direct part  \eqref{direct1}.
Another proof of Hiai-Petz's theorem \eqref{Hiai-Petz-theorem} was also
given by Hayashi \cite{Hayashi-Hiai-Petz}
using the representation theory of the general linear group on $\H$.
In the original proof of Hiai-Petz's theorem \eqref{Hiai-Petz-theorem},
the pinching $\pinching{\sigma_n}{\rho_n}$
defined in Appendix~\ref{appendix:pinching} played an important role.
Since $\pinching{\sigma_n}{\rho_n}$ commutes with $\sigma_n$,
they are diagonalized simultaneously as follows
\begin{align}
\pinching{\sigma_n}{\rho_n}=\sum_{j}\lambda_{n,j}M_{n,j},\quad
\sigma_n=\sum_{j}\mu_{n,j}M_{n,j}.
\end{align}
Finally $M_n=\{M_{n,j}\}$ was shown to be a PVM that 
attains the quantum relative entropy, i.e.,
\begin{align}
\oneover{n} D\left(\pinching{\sigma_n}{\rho_n}\big\|\sigma_n\right)
= \oneover{n} D_{M_n}(\rho_n\|\sigma_n)
\longrightarrow &D(\rho\|\sigma) \quad (n\rightarrow\infty) .
\label{Hiai-Petz-theorem-pinching}
\end{align}

In order to connect Hiai-Petz's theorem \eqref{Hiai-Petz-theorem}
with the direct part \eqref{direct1} and \eqref{direct2},
we needed to apply Stein's lemma in classical hypothesis testing
so far, as mentioned above,
considering independent and identically distributed (i.i.d.) extensions of
the probability distributions $\bigl\{\Tr[\rho_n M_{n,i}]\bigr\}$
and  $\bigl\{\Tr[\sigma_n M_{n,i}]\bigr\}$.
The purpose of this manuscript is to show a direct proof
of \eqref{direct1} and \eqref{direct2} after \cite{Hayashi-Stein},
based on a key operator inequality,
without using the achievability of
the information quantity \eqref{Hiai-Petz-theorem}
nor i.i.d. extensions of the probability distributions.
As is mentioned by Nagaoka \cite{Nagaoka},
the proof also leads to Hiai-Petz's theorem
\eqref{Hiai-Petz-theorem} consequently.
Here a direct proof of Hiai-Petz's theorem \eqref{Hiai-Petz-theorem}
is also shown using the key operator inequality
as well as a proof by way of
the direct part of Stein's lemma \eqref{direct2}.

\section{The Direct Part of Stein's Lemma}

In the sequel,
let us denote $\pinching{\sigma_n}{\rho_n}$
as $\overline{\rho_n}$ for simplicity,
and let $v(\sigma_n)$ be the number of eigenvalues
of $\sigma_n$ mutually different from others
as defined in Appendix~\ref{appendix:pinching}.
A key operator inequality
\footnote{
Although the way to derive the operator inequality
and the definition of $v(\sigma_n)$
are different from those of \cite{Hayashi-Stein},
it results in the same one as \cite{Hayashi-Stein}
in the case that both of $\rho_n$ and $\sigma_n$
are tensored states.
}
follows from Lemma~\ref{lemma:Hayashi}
in Appendix~\ref{appendix:key-inequality},
which was originally appeared in \cite{Hayashi-Stein}.
\begin{lemma}
\begin{align}
\rho_n \le v(\sigma_n)\,\overline{\rho_n}.
\label{key-inequality}
\end{align}
\end{lemma}
Note that the type counting lemma
(see {\it e.g.} \cite{Cover}, Theorem 12.1.1) provides
\begin{align}
v(\sigma_n)\le (n+1)^d.
\label{type-counting}
\end{align}

Following \cite{Hayashi-Stein},
let us apply the operator monotonicity of
the function $x \longmapsto -x^{-s}\,(0\le s\le 1)$
(see {\it e.g} \cite{Bhatia})
to the key operator inequality \eqref{key-inequality}
so that we have
\begin{align}
\overline{\rho_n}^{\,-s} \le v(\sigma_n)^s \rho_n^{-s}.
\label{h1}
\end{align}
Here, let us define
the projection $\proj{ X > 0 }$
for a Hermitian operator $X= \sum_i x_i E_i$ as
\begin{align}
\proj{ X > 0 } \Def \sum_{i:x_i > 0} E_i.
\end{align}
Now we focus on a test defined
with a real parameter $a$ by
\begin{align}
\overline{S}_n(a) \Def \proj{ \overline{\rho_n} - e^{na}\sigma_n > 0 },
\label{def-test}
\end{align}
which satisfies the following theorem.

\begin{theorem}
For $0\le \forall s\le 1$, we have
\begin{align}
\alpha_n\left(\overline{S}_n(a)\right)
&\le (n+1)^{sd}\, e^{n[as-\psi(s)]},
\label{theorem-alpha} \\
\beta_n\left(\overline{S}_n(a)\right)
&\le e^{-na},
\label{theorem-beta}
\end{align}
where
\begin{align}
\psi(s)\Def -\log\Tr\left[
\rho\,\sigma^{\frac{s}{2}}\rho^{-s}\sigma^{\frac{s}{2}}
\right].
\end{align}
\end{theorem}
\begin{proof}
Since the definition of $\overline{S}_n(a)$ provides
$\left( \overline{\rho_n} - e^{na}\sigma_n \right)
\overline{S}_n(a)\ge 0$,
the upper bound on $\beta_n\left(\overline{S}_n(a)\right)$ is given by
\begin{align}
\Tr\left[\sigma_n \overline{S}_n(a)\right]
\le e^{-na} \Tr\left[ \overline{\rho_n}\, \overline{S}_n(a)\right]
\le e^{-na}.
\end{align}
On the other hand, commutativity of operators
$\overline{\rho_n}$ and $\sigma_n$ leads to
$\overline{S}_n(a)=
\proj{ \overline{\rho_n}^s - e^{nas}\sigma_n^s > 0 }$
for $\forall s\ge 0$,
and hence we have 
$\left(\overline{\rho_n}^s - e^{nas}\sigma_n^s\right)
\left(I_n-\overline{S}_n(a)\right)\le 0$.
Note that $\overline{S}_n(a)$ also commutes with $\sigma_n$.
Therefore, taking the property of the pinching \eqref{pinching2}
in Appendix~\ref{appendix:pinching} into  account,
$\alpha_n\left(\overline{S}_n(a)\right)$ is bounded above as follows
\begin{align}
\Tr\left[\rho_n\left(I_n-\overline{S}_n(a)\right)\right]
&=\Tr\left[\overline{\rho_n}\left(I_n-\overline{S}_n(a)\right)\right]
\nonumber\\
&= \Tr\left[\overline{\rho_n}^{\,1-s} \overline{\rho_n}^{\,s}
\left(I_n-\overline{S}_n(a)\right)\right]
\nonumber\\
&\le e^{nas}\Tr\left[\overline{\rho_n}^{\,1-s}\sigma_n^s
\left(I_n-\overline{S}_n(a)\right)\right]
\nonumber\\
&\le e^{nas}\Tr\left[\overline{\rho_n}^{\,1-s}\sigma_n^s\right].
\label{stein-new-1}
\end{align}
Using \eqref{h1}, \eqref{stein-new-1} is bounded above further as
\begin{align}
\Tr\left[\rho_n\left(I_n-\overline{S}_n(a)\right)\right]
&\le e^{nas} \Tr\left[\rho_n\sigma_n^{\frac{s}{2}}\overline{\rho_n}^{\,-s}
\sigma_n^{\frac{s}{2}}\right]
\nonumber\\
&\le v(\sigma_n)^s e^{nas} \Tr\left[\rho_n \sigma_n^{\frac{s}{2}} \rho_n^{-s}
\sigma_n^{\frac{s}{2}}\right]
\nonumber\\
&\le (n+1)^{sd}\, e^{n[as-\psi(s)]},
\end{align}
where the last inequality follows from \eqref{type-counting}.
\end{proof}

\vspace{2ex}
Observing that $\psi(0)=0$ and $\psi'(0) = D(\rho\|\sigma)$,
we can see that for $\forall a \,< D(\rho\|\sigma)$
there exists $0 \le s \le 1$ such that $as-\psi(s) < 0$.
Therefore $\alpha_n\left(\overline{S}_n(a)\right)$
goes to $0$ by \eqref{theorem-alpha}
with its exponent greater than $\max_{0 \le s \le1}\{ -as+\psi(s) \} > 0$,
which leads to a direct proof of \eqref{direct1} and \eqref{direct2}
combined with \eqref{theorem-beta} as asserted.

\section{Hiai-Petz's Theorem}

As pointed out by Nagaoka \cite{Nagaoka},
the direct part \eqref{direct2} leads to
Hiai-Petz's theorem \eqref{Hiai-Petz-theorem} as follows.
For any test $A_n$,
the monotonicity \cite{Lindblad-CP} \cite{Uhlmann}
of the quantum relative entropy provides
\begin{align}
D(\rho\|\sigma)
&= \oneover{n} D(\rho_n\|\sigma_n) \nonumber \\
&\ge \oneover{n} d(\alpha_n(A_n) || 1-\beta_n(A_n) ) \nonumber \\
&= \oneover{n} \bigl\{
 -h(\alpha_n(A_n)) -\alpha_n(A_n)\log(1-\beta_n(A_n))
 - (1-\alpha_n(A_n)) \log \beta_n(A_n) \bigr\} \nonumber \\
&\ge -\frac{\log 2}{n}  - (1-\alpha_n(A_n))\,\oneover{n}\log\beta_n(A_n),
\label{monotonicity2}
\end{align}
where
\begin{align}
d(p||q) \Def p\log\frac{p}{q} + (1-p)\log\frac{1-p}{1-q},\quad
h(p) \Def - p\log p - (1-p)\log(1-p).
\end{align}
Thus a sequence of test $\{A_n\}$ satisfying \eqref{direct2} yields
\begin{align}
D(\rho\|\sigma)
=\lim_{n\rightarrow\infty}\oneover{n} d(\alpha_n(A_n)||1-\beta_n(A_n)),
\end{align}
which means Hiai-Petz's theorem \eqref{Hiai-Petz-theorem}.

On the other hand,
a direct proof of Hiai-Petz's theorem \eqref{Hiai-Petz-theorem}
is shown as follows.
Let us apply the operator monotonicity of
the function $x\longmapsto\log x$
(see {\it e.g} \cite{Bhatia})
to the key operator inequality \eqref{key-inequality}
so that we have
\begin{align}
\log\rho_n \le \log\overline{\rho_n} + \log v(\sigma_n),
\end{align}
and hence
\begin{align}
D(\rho\|\sigma)
&=\oneover{n}D(\rho_n\|\sigma_n)
\nonumber \\
&=\oneover{n}\left\{
\Tr\left[\rho_n\log\rho_n\right] - \Tr\left[\rho_n\log\sigma_n\right]
\right\}
\nonumber \\
&\le\oneover{n}\left\{
\Tr\left[\rho_n\log\overline{\rho_n}\right]
-\Tr\left[\rho_n\log\sigma_n\right] + \log v(\sigma_n)
\right\}
\nonumber \\
&=\oneover{n}\left\{
\Tr\left[\overline{\rho_n}\log\overline{\rho_n}\right]
-\Tr\left[\overline{\rho_n}\log\sigma_n\right]
\right\} + \oneover{n}\log v(\sigma_n)
\nonumber \\
&\le \oneover{n} D\left( \overline{\rho_n} \bigm\| \sigma_n \right)
+\frac{d}{n} \log (n+1),
\end{align}
where the last inequality follows from \eqref{type-counting}.
Combined with \eqref{monotonicity1},
the above inequality leads to \eqref{Hiai-Petz-theorem-pinching}.

\section{Concluding Remarks}

We have shown a new proof of the direct part of
Stein's lemma in quantum hypothesis testing
without using Hiai-Petz's theorem,
based on a key operator inequality satisfied
by a density operator and its pinching.
Compared with \cite{Hayashi-Stein},
the proof is simple and leads to the exponential convergence of
the error probability of the first kind.
The operator inequality has been also used
to show a new proof of Hiai-Petz's theorem.

The original proof of Hiai-Petz's theorem was
drawn from the joint convexity of the quantum relative entropy
and some of its properties.
The joint convexity is shown by applying
the operator convexity of the function $-\log x$ 
to the relative modular operator, today.
On the other hand,
our proof has been completed with less mathematical preparations
because it is given only by
the operator monotonicity of the function $\log x$.

\appendices

\section{Definition of the Pinching}
\label{appendix:pinching}

In this appendix, we summarize the definition of the pinching
and some of its properties.
Given an operator $A\in\L(\H)$,
let $A=\sum_{i=1}^{v(A)} a_iE_i$ be its spectral decomposition,
where $v(A)$ is the number of eigenvalues of $A$
mutually different from others, and each $E_i$ is the projection
corresponding to an eigenvalue $a_i$.
The following map defined by using the PVM $E=\{E_i\}_{i=1}^{v(A)}$
is called the pinching:
\begin{align}
\E_{A}: B \in\L(\H) \longmapsto \E_{A}(B)
\Def\sum_{i=1}^{v(A)} E_i B E_i \in\L(\H).
\label{pinching1}
\end{align}
The operator $\E_{A}(B)$ is also called the pinching when no confusion
is likely to arise, and it is sometimes denoted as $\E_{E}(B)$.
It should be noted here that
$\E_{A}(B)$ commutes with $A$ and we have
\begin{align}
\Tr[BC] = \Tr\left[\E_{A}(B)C\right]
\label{pinching2}
\end{align}
for any operator $C\in\L(\H)$ commuting with $A$.

\section{Proof of the Key Operator Inequality}
\label{appendix:key-inequality}

The following lemma was appeared in \cite{Hayashi-Stein},
and played an important role in this manuscript.
We show another proof later as well as
the original proof for readers' convenience.
\begin{lemma}[Hayashi \cite{Hayashi-Stein}]
Given a PVM $M=\{M_i\}_{i=1}^{v(M)}$ on $\H$,
we have for $\forall\rho\in\S(\H)$
\begin{align}
\rho \le v(M)\E_M(\rho),
\end{align}
where $\E_M(\rho)$ is the pinching defined
in Appendix~\ref{appendix:pinching}.
\label{lemma:Hayashi}
\end{lemma}
\begin{proof}
It is sufficient to prove the operator inequality for  a pure state
$|\phi\rangle\langle\phi| \in\S(\H)$ as follows.
For $\forall\psi\in\H$, we have
\begin{align}
\bigl\langle\psi\bigl|
\bigl(\, v(M)\E_M(|\phi\rangle\langle\phi|)-|\phi\rangle\langle\phi| \,\bigr)
\bigr|\psi\bigr\rangle
= v(M) \sum_{i=1}^{v(M)} \bigl| \langle\psi| M_i |\phi\rangle \bigr|^2
-\Biggl| \sum_{i=1}^{v(M)} \langle\psi| M_i |\phi\rangle \Biggr|^2
\ge 0,
\end{align}
which follows from Schwarz's inequality about complex vectors
$\bigl(\,\langle\psi| M_i |\phi\rangle\,\bigr)_{i=1}^{v(M)}$
and $\bigl( 1 \bigr)_{i=1}^{v(M)}$.
\end{proof}

\vspace{2ex}
We can also show another proof of Lemma~\ref{lemma:Hayashi}
by using the following operator convexity.
\begin{lemma}
Given a nonnegative operator $A\in\L(\H)$,
the following map is operator convex.
\begin{align}
f_A: X\in\L(\H) \longmapsto X^*AX\in\L(\H).
\end{align}
In other words, we have
\begin{align}
f_A(tX+(1-t)Y)\le tf_A(X)+(1-t)f_A(Y)
\end{align}
for $\forall X,Y\in\L(\H)$ and $0\le \forall t\le 1$.
\label{lemma:convexity}
\end{lemma}
\begin{proof}
The assertion is shown by a direct calculation as follows
\begin{align}
&tf_A(X)+(1-t)f_A(Y) - f_A(tX+(1-t)Y) \nonumber \\
&= tX^*AX + (1-t)Y^*AY - [tX+(1-t)Y]^*A\,[tX+(1-t)Y] \nonumber \\
&= t(1-t) [X^*AX -X^*AY-Y^*AX+Y^*AY] \nonumber \\
&= t(1-t) (X-Y)^*A\,(X-Y) \nonumber \\
&\ge 0.
\end{align}
\end{proof}

\vspace{2ex}
\noindent
Now Lemma \ref{lemma:Hayashi} is verified
by using Lemma \ref{lemma:convexity} as follows
\begin{align}
\oneover{v(M)^2}\rho
&= \Biggl( \oneover{v(M)}\sum_{i=1}^{v(M)}M_i \Biggr)
 \,\rho\, \Biggl( \oneover{v(M)}\sum_{i=1}^{v(M)}M_i \Biggr) \nonumber \\
&\le \oneover{v(M)} \sum_{i=1}^{v(M)} M_i \rho M_i \nonumber \\
&= \oneover{v(M)}\E_{M}(\rho).
\end{align}

\section*{Acknowledgment}

The authors are grateful to Prof.~Hiroshi~Nagaoka.
He encouraged them to show a simple proof of the direct part
of Stein's lemma in quantum hypothesis testing
pointing out that the proof leads to Hiai-Petz's theorem.

This research was partially supported by
the Ministry of Education, Culture, Sports, Science, and Technology
Grant-in-Aid for Encouragement of Young Scientists, 13750058, 2001.


\end{document}